# The Role of Glia Cells in the Basic Algorithm of the Formation of Memory in the Brain

Charles Ross and Shirley Redpath

## Abstract

*We build on recent progress in understanding the role of glia cells in building the initial neural networks in the foetal brain. This has led us to make three significant postulates. Short term memory results from glia cells forming speculative links directly and solely as a result of neural activity generated by life experiences. These temporary 'glia bridges' create long term memory by stimulating the growth of axons, dendrites and synapses and providing the pathways enabling permanent neural structures to be created. Problem-solving, idea creation and memory maintenance result from this fundamental algorithm for how the brain generates new links.*

**Argument**

The development of the brain falls into two main phases. The basic framework of some billion neural networks is constructed from conception to birth and then on to maturity under the general stimulus and control of our inherited DNA. From the moment the nascent brain begins to receive signals from external stimuli, it then begins to grow trillions of additional links, which continue to develop as a result of our experiences and learning throughout our lives.

This paper sets out to discuss six key questions about the formation of these links:-

1. What stimulates neuron networks to initiate new links or circuits?
2. What is the physical process involved when an individual neuron grows a new dendrite or axon filament?
3. How does that filament know where to go?
4. How does it know how to get there?
5. What is the relationship between short term, expendable memories and long term permanent memories?
6. How is it that neural networks appear to migrate over time?

The first neural structures appear within 14 days of conception, a heart beat within three weeks and recognisable continuous electrical activity can be measured in early neural nets on the proto cortex within 12 weeks.[1] By birth, approximately one billion neural networks will directly or indirectly connect every cell in the body to the brain, enabling that organ to control the behaviour of the entire body as one cooperative, coordinated whole. In particular, neuron networks connect the sense organs to the brain (input system), and the brain to the muscles and other organs (output system). This structure provides the process that enables humans to move muscles in response to received information from the senses. That general ability is refined into specific motor skills. Similarly this framework enables the interpretation of sounds, images, tastes, smells and touch sensations. It also enables the creation of new sounds and images.

By combining the basic inherited framework with the dynamic ability to learn and refine skills, Nature has equipped humans with a powerful force for evolutionary development. For instance, newborn humans are equipped with the propensity to make and hear the sounds of language, but have to learn to articulate and decode every single word of the language used by the people around them.[2] In other words; humans are born with a brain engineered by DNA to do almost nothing. Only with the structures for learning in place are humans able to do anything. Throughout life, this general enabling framework, donated by DNA is populated with the specific tailored neural linkages to execute every physical task, learn to solve problems and create new ideas.

How does this occur? Research is amassing considerable knowledge about the foetal development phase which is guiding us towards an understanding of the later learning phases. Neurons appear to be built in manufacturing sites and then migrate to their destinations. Like 'stem cells', neurons initially have the potential to develop into any type and fulfil any role.[3] Each neuron possesses the same standard set of DNA, so the specific type of neural cell it develops into is dependent on how the genes are expressed. Single neurons do not build networks on their own. They appear to work in teams. Before the neurons start their journeys a matrix of glia cells provide scaffolding for the construction of the networks.[4]

Various neurotransmitters broadcast chemical messengers which stimulate the development of different types of axons and dendrites. Neural nuclei start to manufacture strings of protein cells forming teams of axon and dendrite filaments which slide along the glia scaffolding up the chemical gradient towards target destinations. Both neurons and glia use cell and substrate adhesion molecules to assist in these activities. When one member of the team is safely anchored into place with a secure synaptic connection, the remainder of the supporting team dissolves. All rely directly or indirectly on the precise instructions in the DNA.

From the moment of birth humans need something more than this inherited blueprint. We have to respond to the environment in which we find ourselves. DNA itself cannot build even one single neural network to help us learn to walk or speak one word of the language of our parents and peers. Some other additional process and agency is at work.

Nature is a notoriously lazy inventor, preferring to utilise existing processes over developing new ones. It seems likely, therefore, that given the crucial role of glia cells in the foetal stage of brain development, they play an equally important role in populating the brain with the neural networks of experience. Learning is about building memories, and memories are sets of instructions for responding to the experienced environment. For instance in order to say a word and experience the feedback it receives from the outside world, the brain has to grow specific neural links to coordinate the muscles of the lungs, vocal chords, mouth, tongue and lips to create the exact sound. We now have a number of clues as to how that might occur. A study of *Aplysia* gives two rules for this [6].

Firstly, the repetition of patterns of stimulation causes neurons to grow dendrite and axon filament and synapses. These new structures modify existing networks and build new ones, both to differentiate incoming patterns, and to initiate the sets of instructions which control the muscle activity that implements responses. In short, the arrival of information drives the system. Secondly, whenever neural networks are active concurrently, new links form between them.

It has been suggested that connections between neurons that fire at the same time will be strengthened[7]. CREB, one of a series of complicated transcription factor proteins inside the nucleus, plays a key role in memory development by activating the genes in the nucleus to manufacture the proteins needed to grow axons and dendrites.[8]

Glia cells also seem to play a role. Neurons grow very few synapses if no glia cells are present[9], and a particular type of glia cell, a Schwann cell, plays a key role in helping damaged axons that have become detached from their synaptic connection to find their way back, by providing them with a path to follow[10]. It is also known that glia cells pack around the sites of potential synapses, possibly assisting in their formation.[11]

Another type of glia cells, the Astrocyte cells; appear to prompt ATP to activate communication along a string of glia cells[12]. While neurons communicate through a series of established links, much like telephone wires, glia cells broadcast their messages by transmitting neurotransmitters which can be picked up by receptors on nearby neurons, enabling communication between networks before a direct link is established.

**Discussion**

From consideration of the results outlined above, we seek to construct what seems like a plausible hypothesis.

We have noted that the brain seems to be driven almost entirely by the information that it receives – by experience. When a sense or other organ, or another neuron is activated by a perceived stimulus it sends a stream of electrical signals along the dendrite and axon filaments, activating the neural networks of the brain. These electrical signals, or action potentials, generate weak temporary electro-magnetic fields around the filaments for the

length of time that they are active. In addition they stimulate the output of neurotransmitters
If the filaments are close together and their electromagnetic fields overlap, we postulate that the combined magnetism of these two overlapping fields is sufficient to attract some of the free glia protein cells in the extracellular fluid that bathes the brain to move a few Ångstrom and form a temporary link. This 'bridge' is able to carry an electrical message between the active neuron filaments, thereby connecting the two networks together. Alternatively, if the two neural networks are some distance apart they may output neurotransmitters that attract a longer string of glia to carry out the same task.

Human brains grow millions of new neural links every second[5]. This hypothesis may provide the mechanism. These speculative bridges or linking circuits form continuously and automatically whenever pairs of neurons are routinely activated in response to the activity in the environment. The vast majority of these temporary links will not be used again and so will dissolve. Those that remain provide networks of temporary dendrite and axon surrogates that provide short term memory.

There seem to be a number of mechanisms through which these temporary bridges can form. There are several candidate glia cells and, indeed, there may be different systems involved to create different links. The observed activity of both Schwann glia cells and Astrocyte glia cells in the stimulation and repair of neural network functions has been noted above. In the visual area of the brain, the tectum emits the hormone neurotropin, which appears to help in the formation of axons. Neurotransmitters might also play a role in linking the glia together to form chains – a succession of mini synapses perhaps between a string of glia cells[12].

It seems likely that the formation of these temporary links is strongly modulated by the ambient chemical mix in the brain - the current emotional state. In a state of high arousal or elevated concentration, the brain is flooded with steroids and hormones, substantially increasing the chances of a new link being formed and strengthened [13].

Thus we have outlined an algorithm for the formation of transitory speculative short term 'memories' caused automatically by the activities of the neural networks in response to our experiences. Now we will address how these short term memories are replaced by long term protein axons, dendrites, and synapses.

The DNA in the nucleus of each neuron is responsible both for building its axons and dendrites in their original formation and for their subsequent control. When additional new pathways of electrical activity - 'glia bridges' - come into operation, they affect the way the nucleus controls additional information and the destinations to which it despatches response signals. As the DNA accommodates these new facilities, it seems logical that continuous activity along any of these temporary circuits will cause it to initiate the building of more permanent structures. Making permanent memory involves stimulating cells to produce proteins and so it seems plausible that the electrical activity generated by repeated use of the new temporary 'glia bridge' link activates the genes in the DNA.

Axon activity can influence the read out of genes in a glial cell and thus influence its behaviour[14]. It is only one further step to suggest that axon and dendrite activity can influence the behaviour of genes in a neuron nucleus.

This highlights a persuasive and plausible process for the creation of the neural networks that encapsulate memories. The processes of foetal neurogenesis are repeated after birth. The difference is that pre-natal networks are generated by the DNA blueprint and operating instructions, while the post-natal networks are generated automatically as a result of the physical information flow generated through our experiences.

## Summary

We are now in a position to offer answers to the questions outlined at the beginning of this paper.

The electrical activity generated by active neural networks stimulates the initiation of speculative temporary glia links between them. These new links are modulated by the ambient chemical mix (emotional state) and strengthened by usage. Active temporary links stimulate the nuclei to replace them with permanent axons and dendrites filaments and synapses. Thanks to the existence of the temporary glia bridges, the new filaments know where they are going, and can follow these pathways to get there. As with the foetal process, it seems likely that neural activity causes many initial links to be formed, but only a few will eventually be established as new permanent links. Through this system, information arriving in the brain automatically stimulates the formation of speculative temporary links that can lead to their replacement by permanent neural memory paths and so, for instance, build the motor circuits to execute activities and so build physical abilities; and connect words to images and other facts and so create knowledge – the basis of learning.

Question six is more profound. Neurons of themselves have no knowledge of past or future. Every time neural networks are stimulated they automatically generate temporary speculative links, unaware of whether this is the first event or a well established activity. Thus, every time the neural circuit of a memory is accessed, we create a new experience which carries potential variations on the original: various ways of access, various new structures, and various outputs. If the initial structure is used continuously, these variations will develop as associated networks in other areas of the brain, providing valuable redundancy and plasticity to cope with damage or wear. This explains how brain scans appear to show well established activities occurring in more than one location in the brain.

The brain continues to build stem cell like neurons well after birth and sometimes these get incorporated into new networks by helping to replace temporary 'glia bridges' that are very active[3].

**Conclusion**

In this paper we have presented an algorithm that provides a very useful framework to illuminate many brain functions. In particular it provides an algorithm to explain how humans can learn new tasks such as swimming, balancing a bicycle, driving a car, or reciting poetry. Though initially unable to coordinate the many bodily responses needed for competent execution of the skill and thus involving every available facility, repetition enables the brain to construct the minimum efficient neural circuits needed to become expert.[15]

The concept of temporary bridges automatically linking up active networks could also help explain how humans are able to come up with a novel solution to a problem after puzzling over it for some time without success. At some point in the deliberations, two previously unrelated networks grow a speculative temporary link which generates a flash of creativity.

We offer two further speculations for consideration. First, if the development of successful temporary 'bridges' causes the DNA in the nucleus to grow a new axon or dendrite, it is possible that this activity may also reciprocally modify the DNA, or at least the expression of the DNA we inherit. As the DNA in a cell is replicated, it stands to reason that any modification, in the form of new axon and dendrite growth, must be contained in the replication if the brain function dependent on the new filaments is not to be lost. It follows that copies of the modified DNA are passed out of the neuron so that if the neuron is damaged or wears out, the blueprint is available to build a replacement neuron that includes everything that had been created – learned - after birth, and so incorporating *extragenetic* experiences and learning.[16]

Second, if modified DNA circulates in the body it could influence the DNA laid down in the sperm and ova, so potentially modifying the genes of subsequent generations. Thus, we would have a process whereby every generation is able to benefit from all the accumulated experiences of their forebears. Here we have a possible algorithm for *Transgenerational Epigenetic Inheritance*[17]

Charles.T. Ross (Hon)FBCS, CITP, FIMIS, FIAP.
Charles.Ross@BrainMindForum.co.uk

Shirley. F. Redpath, MA, MBA.
Shirley.Redpath@BrainMindForum.co.uk

These hypotheses, ideas, theories, conjectures and their implications are set out in greater detail in *Biological Systems of the Brain: Unlocking the Secrets of Consciousness*, by Charles Ross & Shirley Redpath.
Brain Mind Forum
ISBN 978 1848760 004  http://www.troubador.co.uk/book_info.asp?bookid=671